\documentclass[showpacs,pra,onecolumn,superscriptaddress]{revtex4}
\usepackage{amsmath}
\usepackage{graphicx}
\usepackage{dcolumn}
\usepackage{bm}
\begin{document}
\title{Entanglement and symmetry effects in the transition to the Schr\"{o}dinger
cat regime}

\author{Ferdinando  de Pasquale}\email{ferdinando.depasquale@roma1.infn.it}
\affiliation{Dipartimento di Fisica, Universit\`{a} di Roma La
Sapienza, Piazzale A. Moro 2, 00185 Roma, Italy}
\affiliation{CNR-INFM Center for Statistical Mechanics and
Complexity}
\author{Gian Luca Giorgi}\email{gianluca@ifisc.uib-csic.es}
\affiliation{Institute for Cross-Disciplinary Physics and Complex Systems, IFISC (CSIC-UIB),
Campus Universitat Illes Balears, E-07122 Palma de Mallorca, Spain}
\author{Marco Zannetti}\email{zannetti@sa.infn.it}
\affiliation{Dipartimento di Matematica e Informatica, Universit\`{a} di Salerno, via Ponte don Melillo, 84084 Fisciano (SA), Italy}


\begin{abstract}
We study two-spin entanglement and order
parameter fluctuations as a function of the system size in the XY model in a
transverse field and in the isotropic XXX model. Both models are
characterized by the occurrence of ground state degeneracy also
when systems of finite size are considered. This is always true
for the XXX model, but only at the factorizing field for the XY
model. We study the size dependence of symmetric states, which, in
the presence of degeneracy, can be expanded as a linear
combination of broken symmetry states. We show that, while the XY
model looses its quantum superposition content exponentially with
the size $N$, a decrease of the order of $1/N$ is observed when
the XXX model is considered. The emergence of two qualitatively
different regimes is directly related to the difference in the
symmetry of the models.
\end{abstract}
\maketitle

\section{Introduction}

The transition between the microscopic and macroscopic worlds is a
fundamental issue in quantum theory both from the point of view of
foundations of physics and of the application to quantum computation \cite{giulini,zurek}.

Spontaneous symmetry breaking (SSB) indicates a situation where,
given a symmetry of the Hamiltonian, there are eigenstates which
are not invariant under the action of this symmetry, unless a term
is added which explicitly breaks the symmetry. Usually, when the
control parameter reaches a critical value, the lowest energy
eigenstate keeping the Hamiltonian symmetry is no longer stable in
the presence of infinitely small perturbations, and new stable
solutions appear which are not symmetric. SSB leads naturally to a
degenerate manifold of ground states.

Symmetry breaking usually occurs in the thermodynamic limit,
when superselection destroys quantum coherence. Important exceptions are the
XXX Heisenberg model and the XY model in a transverse magnetic field at the
particular value of the field where ground state factorization occurs \cite{kurmann}. In these cases, ground state degeneracy occurs for any size of the system,
and it is therefore possible to explicitly study the transition from quantum
to classical behavior. How entanglement is affected, in the thermodynamic limit, by the presence of a term which explicitly breaks the symmetry, ha been discussed by Sylju\aa{}sen \cite{syl} and by Osterloh \textit{et al.} \cite{osterloh}. Here, we face the problem from another point of view, starting from small systems, and then increasing the size until the thermodynamic limit is reached. 

The XY model in a transverse field has been introduced in the
early sixties and solved by Katsura \cite{katsura}, by means of
the Jordan-Wigner transformation, formerly introduced by Lieb,
Schultz, and Mattis \cite{lieb}. Subsequently, the correlation
functions were investigated in great detail by Barouch and MacCoy
\cite{barouch}, who found the existence of a second critical value
of the transverse field separating qualitatively different
behaviors of the correlation functions. Later on, Kurmann, Thomas and Muller \cite{kurmann}
discovered ground state factorization for a large class of spin
models. In the particular case of the XY model, the field at which
factorization occurs is exactly the critical field of Barouch and
MacCoy. Recent interest has been devoted to the study of
entanglement properties of many-body systems undergoing a quantum
phase transition \cite {osborne,fazio,vidal,vedral}. As shown in
Ref. \cite{amico}, the critical point turns out to separate two
regions with qualitatively different bipartite entanglement. It has been shown in Ref. \cite{baroni} that, in the vicinity of the factorizing field, the range of concurrence diverges, and that such divergence
corresponds the appearance of a characteristic length scale in the system.
Recently, the conditions for the existence of the factorizing field for models with long-range interaction have been determined \cite{giampaolo}, and the study of this effect has been extended to dimerized chains \cite{gianluca}.

The two systems we wish to investigate belong to different classes of
symmetry. While the XXX Heisenberg model has the SU(2) continuous symmetry,
i.e. the Hamiltonian commutes with the total spin along any possible direction, the XY
model is invariant under parity transformations and possesses the discrete $Z_{2}$
symmetry.

It is commonly accepted that purely quantum effects are not
observable on the macroscopic scale, except for superconductivity,
superfluidity. On the other hand, quantifying entanglement
(perhaps the most genuine manifestation of quantum properties) as
a function of the system size represents a fundamental issue
\cite{vednat}. Here, we wish to investigate in detail the two-spin
entanglement dependence on the total number of spins for these
models. In particular, we shall derive the difference in the size
effects due to the difference in the system symmetry. To be more
specific, in the case of a discrete symmetry there is an
exponential entanglement decrease, while with a continuous
symmetry entanglement shrinks linearly with the growth of the
system size.

Since the XY model has been studied through the last four decades, and
results are scattered over a vast literature, for convenience we shall give here a brief survey
of the main results, with the primary aim of focusing on the existence of the
factorizing field and its independence from the system size.

The paper is organized as follows. In Sec. \ref{maro} the XY model
in a transverse field is discussed. We give special emphasis to
the finite size solution with the scope of enlightening the
emergence of the factorizing field as a size-independent
degeneracy point. Furthermore, by means of the finite size
picture, we are able to explain in a simple way the appearance of
spontaneous symmetry breaking in the thermodynamic limit. In Sec.
\ref {XXX} we describe briefly the structure of the ground state
for the isotropic Heisenberg (XXX) model. Even if, for any finite
number of spins the ground state manifold has finite dimension, an
over-complete set of states can be introduced that allows to study
the microscopic-to-macroscopic transition. In Sec. \ref{2sp} we
derive the value of the concurrence for pairs of spins and the
order parameter fluctuation in a superposition state as a function
of the size systems both for the XY model and the XXX model.
Finally, in Sec. \ref{disc}, results are discussed. In particular,
we will focus on the influence of the symmetry in the different
behaviors.

\section{XY Model\label{maro}}

Let us consider a chain of $N$ spins
\begin{equation}
H=\sum_{l}\left[ J\frac{\left( 1+\gamma \right) }{2}\sigma _{l}^{x}\sigma
_{l+1}^{x}+J\frac{\left( 1-\gamma \right) }{2}\sigma _{l}^{y}\sigma
_{l+1}^{y}+h\sigma _{l}^{z}\right] ,
\end{equation}
where $\sigma ^{\epsilon }$ are the three Pauli matrices $\left( \epsilon
=x,y,z\right) $, and periodic boundary conditions ($\sigma _{N+1}^{\epsilon
}=\sigma _{1}^{\epsilon }$) are assumed. In the following we will assume $J=-1
$ (ferromagnetic coupling). The above Hamiltonian is invariant under the $%
Z_{2}$ group of the rotations by $\pi $ about the $z$ axis, since
it commutes with the parity operator $P=\prod_{l}\sigma _{l}^{z}$.
Due to this symmetry, eigenstates are classified depending on
parity eigenvalue. This system is known to undergo a quantum phase
transition at the critical point $h_{c}=1$. Below this value, in
the thermodynamic limit, spontaneous magnetization along the $x$
axis appears.

Since the work of Ref. \cite{lieb}, the Jordan-Wigner transformation,
defined through $\sigma _{l}^{z}=1-2c_{l}^{\dagger }c_{l}$, $\sigma
_{l}^{+}=\prod_{j<l}\left( 1-2c_{l}^{\dagger }c_{l}\right) c_{l}$, $\sigma
_{l}^{-}=\prod_{j<l}\left( 1-2c_{l}^{\dagger }c_{l}\right) c_{l}^{\dagger }$
is introduced to map spins in spinless fermions. The transformed Hamiltonian
is $H=H_{0}-PH_{1}$ with
\begin{eqnarray}
H_{0} &=&-\sum_{l=1}^{N-1}\left[ \left( c_{l}^{\dagger
}c_{l+1}-c_{l}c_{l+1}^{\dagger }\right) +\gamma \left( c_{l}^{\dagger
}c_{l+1}^{\dagger }-c_{l}c_{l+1}\right) -h\left( 1-2c_{l}^{\dagger
}c_{l}\right) \right] , \\
H_{1} &=&-\left[ \left( c_{N}^{\dagger }c_{1}-c_{N}c_{1}^{\dagger }\right)
+\gamma \left( c_{N}^{\dagger }c_{1}^{\dagger }-c_{N}c_{1}\right) \right] .
\end{eqnarray}
Since $\left[ H,P\right] =0$, all eigenstates of $H$\ have
definite parity, and we can proceed to a separate diagonalization
of $H$ in the two subspaces labelled by to $P=\pm 1$. Then, the
complete set of eigenvectors of $H$ will be given by the even
eigenstates of\ $H^{+}=H_{0}-H_{1}$ and the odd eigenstates of
$H^{-}=H_{0}+H_{1}$. Both for $H^{+}$\ and $H^{-}$ the
diagonalization is obtained by first carrying out the space
Fourier transform
\begin{equation}
c_{k}=\frac{1}{\sqrt{N}}\sum_{k}e^{-i\frac{2\pi }{N}kl}c_{l},
\end{equation}
where $k=0,1,\ldots ,N-1$ in $H^{-}$, and $k=1/2,3/2,\ldots ,N-1/2$ in $H^{+}
$, and then making the Bogoliubov transformation
\begin{equation}
c_{k}=\cos \vartheta _{k}\eta _{k}+i\sin \vartheta _{k}\eta
_{-k}^{\dagger },
\end{equation}
with $\vartheta _{k}=-\vartheta _{-k}$. Here, $\eta _{-k}$ stands for $\eta _{N-k}$.
The diagonalization condition implies for $\vartheta _{k}$
%\begin{mathletters}
\begin{equation}
\tan 2\vartheta _{k}=-\frac{\gamma \sin k}{h-\cos k}.
\end{equation}
Eventually, we end up with the quasi-particle Hamiltonians
%\end{mathletters}
\begin{eqnarray}
H^{+} &=&\sum_{k=1/2}^{N-1/2}\Lambda _{k}\left( \eta _{k}^{\dagger }\eta
_{k}-\frac{1}{2}\right) , \\
H^{-} &=&\sum_{k=0}^{N-1}\Lambda _{k}\left( \eta _{k}^{\dagger }\eta _{k}-%
\frac{1}{2}\right) ,
\end{eqnarray}
where the eigenvalues are given by
\begin{equation}
\Lambda_{k}=2\sqrt{\left(h-\cos \frac{2\pi}{N}k\right)^{2}+\gamma^{2}\sin^{2}\frac{2\pi}{N}k}.
\end{equation}
The ground states of $H^{+}$ and $H^{-}$
are the corresponding vacuum states with eigenvalues
$E_{0}^{+}=-\sum_{k=1/2}^{N-1/2}\Lambda _{k}$ and $%
E_{0}^{-}=-\sum_{k=0}^{N-1}\Lambda _{k}$.

The vacuum in the generic $k$ mode is determined by $\eta _{k}\left| 0^{\pm }\right\rangle =0$.\
While for every $k\neq 0$ the Bogoliubov vacuum corresponds to an even state
(the absence of quasi-particles implies zero or two particles), the mode $k=0
$ plays a special role. In fact, the correspondent Bogoliubov transformation
reads
\begin{equation}
\eta _{0}=\frac{1}{2}\left( 1+\frac{h-1}{\left| h-1\right| }\right) c_{0}+%
\frac{i}{2}\left( 1-\frac{h-1}{\left| h-1\right| }\right)
c_{0}^{\dagger },
\end{equation}
with the important consequence that the quasi-particle vacuum corresponds to a
zero-particle state for $h<1$ and to one-particle state for $h>1$. The
presence or the absence of the particle in the $k=0$ mode changes the parity
of the state. Thus, for $h>1$, the vacuum of $H^{-}$, because of its
symmetry, does not belong to the set of eigenstates of $H$, while for $h<1$
it becomes an eigenstate of physical interest. Above the $h=1$, the odd
state of lowest energy is obtained by adding one quasi-particle
corresponding to the bottom of the energy band with energy $\Lambda _{\min
}=2\left( h-1\right) $. This energy gap prevents the degeneracy even in the
thermodynamic limit.

\subsection{Quantum phase transition and ground state factorization}

The change of symmetry of the vacuum of H$^{-}$ is the very cause of the
phase transition in the thermodynamic limit. Indeed, on the macroscopic
scale the sum over $k$ becomes an integral yielding $E_{0}^{+}=E_{0}^{-}$. Then,
below the critical point $h_{C}=1$ the odd and the even lowest eigenstates
are degenerate, and the Hamiltonian symmetry is spontaneously broken, while,
for $h>h_{C}$, \ due to the existence of the energy gap $\Lambda _{\min }$,
the ground state keeps its parity (even). For $h<h_{C}$, because of
superselection rules, the system is necessarily found in symmetry-broken
states.

As pointed out in Ref. \cite{barouch}, below the critical point there are
two different regions where two-body correlation functions can decrease
monotonically or oscillate as a function of the spin distance, depending on
the Hamiltonian parameters. These regions are separate, in the $\{h,\gamma \}
$ diagram, by the set of points satisfying $h_{F}^{2}+\gamma ^{2}=1$.
More recently, it has been shown that on this border line the ground state
factorizes \cite{verrucchi}, i.e. it can be written as $\left| \Psi
_{F}^{\pm }\right\rangle =\otimes _{l}\left| \Psi _{F,l}^{\pm }\right\rangle
$, with $\left| \Psi _{F,l}^{\pm }\right\rangle =\left( \cos \alpha \left|
\uparrow _{l}\right\rangle \pm \sin \alpha \left| \downarrow
_{l}\right\rangle \right) $, where $\cos 2\alpha =\left[ \left( 1-\gamma
\right) /\left( 1+\gamma \right) \right] ^{1/2}$.

\begin{figure}
  \includegraphics[width=.5\textwidth,height=40mm]{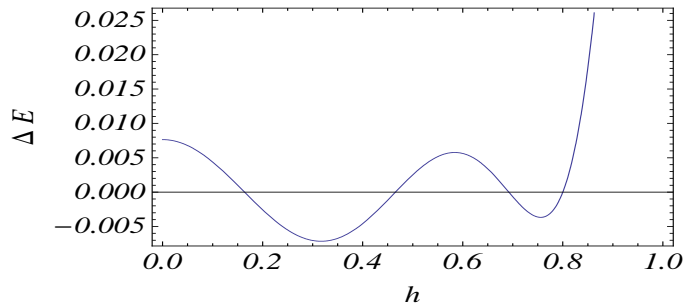}
\caption{Difference of energy between the lowest odd and even
eigenvalues of $H$ as a function of the transverse field for a
8-spin chain. Given the anisotropy amplitude $\gamma=0.6$, the
factorizing field is $h_F=0.8$. As predicted, we observe $N/2$
level crossing point, the last of them being $h_F$.}
\label{energie}
\end{figure}

The existence of the factorizing field, originally derived by
requiring only size-independent degeneracy between the lowest odd
and even eigenvalues \cite {kurmann}, can be studied within the
general solution of the model. By analyzing lowest odd and even
eigenvalues of $H$ in the symmetry broken region for finite $N$ as
a function of the transverse field, we observe a series of $N/2$
level crossings for $h=h_{i}$ (see Fig. \ref{energie}). In
correspondence of each $h_{i}$ the ground state changes its
symmetry. The existence of such points has been discussed in Ref.
\cite{hoeger} and more recently in Ref. \cite{rossignoli}, and is
responsible for the magnetization jumps reported in Ref.
\cite{puga}. In the thermodynamic limit, this kind of structure
implies two different symmetry breaking mechanisms. For
$0<h<h_{F}$, as $N\rightarrow \infty $, the set $\left\{
h_{i}\right\} $ of the degeneracy points becomes a denumerable
infinity, while for $h_{F}<h<1$ there is a the usual symmetry
breaking due to the vanishing of the gap.
%This is the microscopic mechanism responsible for dissimilar
%two-body correlation functions.({\bf metterei quest'ultima frase
%in forma pi\`u dubitative, tipo:
An interesting problem would be to check whether this is the
microscopic mechanism responsible for the qualitative change in
the behavior of the correlation functions above and below $h_{F}$.

% 
% \begin{figure}
%   % Requires \usepackage{graphicx}
%   \includegraphics{energie.eps}\\
%   \caption{Difference of energy between the lowest odd and even
% eigenvalues of $H$ as a function of the transverse field for a
% 8-spin chain. Given the anisotropy amplitude $\gamma=0.6$, the
% factorizing field is $h_F=0.8$. As predicted, we observe $N/2$
% level crossing point, the last of them being $h_F$.}
% \label{energie}
% \end{figure}

While spontaneous symmetry breaking arises only for $N\rightarrow
\infty $, it can be seen from the previous analysis (see also Ref.
\cite{rossignoli}) that at the factorizing point $h_{F}$
degeneracy appears for any $N$. It is simple to show that
$E_{0}^{+}\left( h_{F}\right) =E_{0}^{-}\left( h_{F}\right) $
holds for any $N$, while the positions of all the other level
crossing points $h_{i}$ change with $N$. Then, at the special
field $h_{F}$, the Hamiltonian symmetry is broken independently of
the system size, and any linear superposition of the two symmetric
eigenstates $\left| \alpha ^{\pm }\right\rangle $ ($\left| \alpha
^{+}\right\rangle $ for the even eigenstate and $\left| \alpha
^{-}\right\rangle $ for the odd eigenstate), is a possible
eigenstate. Obviously, each one of the factorized states can be
expressed as a linear combination of the two\thinspace symmetric
eigenstates
\begin{equation}
\left| \Psi _{F}^{\pm }\right\rangle =u_{+}\left| \alpha
^{+}\right\rangle \pm u_{-}\left| \alpha ^{-}\right\rangle ,
\label{ftoa}
\end{equation}
with $u_{\pm }=\left[ \left( 1+\cos ^{N}2\alpha \right) /2\right] ^{1/2}$.
Notice that, for finite size systems, $\left| \Psi _{F}^{+}\right\rangle $
and $\left| \Psi _{F}^{-}\right\rangle $ are not orthogonal, while $\
\left\langle \Psi _{F}^{+}|\Psi _{F}^{-}\right\rangle =0$ in the
thermodynamic limit.

\section{XXX model\label{XXX}}

%
%({\bf controllare se cos\`i il paragrafo seguente va bene})

The homogeneous (ferromagnetic) Heisenberg model is defined by the
Hamiltonian
\begin{equation}
H_{XXX}=-J\sum_{l=0}^{N-1}\left( \sigma _{l}^{x}\sigma _{l+1}^{x}+\sigma
_{l}^{y}\sigma _{l+1}^{y}+\sigma _{l}^{z}\sigma _{l+1}^{z}\right) ,
\end{equation}
with the boundary condition $\sigma _{N}^{\epsilon }=\sigma
_{0}^{\epsilon }$. The model has been solved using the Bethe
ansatz \cite {bethe}. As far as the ground state properties are
concerned, a simple argument can be introduced to show that, for
any number of spins, any factorized
state $\left| \Phi \left( \theta ,\phi \right) \right\rangle =\otimes _{l}%
\left[ \cos \theta \left| \uparrow \right\rangle +\exp \left( i\phi \right)
\sin \theta \left| \downarrow \right\rangle \right] $ minimizes the energy.
Given the invariance of $H_{XXX}$ with respect to rotations of arbitrary
amplitude $\beta $ around any direction $\hat{n}$ $\mathcal{R}\left( \beta ,%
\hat{n}\right) =\prod_{l}\exp \left[ i\beta \vec{\sigma}_{l}\cdot \hat{n}%
\right] $, due to $\left[ H_{XXX},\mathcal{R}\left(
\beta ,\hat{n}\right) \right] =0$, we shall restrict the attention on the particular state $\left| \Phi
\left( 0,0\right) \right\rangle =\left| \uparrow ,\uparrow ,\ldots ,\uparrow
\right\rangle $. It can be immediately seen that $\left| \Phi \left(
0,0\right) \right\rangle $ belongs to the ground state subspace of any of
two-body terms of $H_{XXX}$, and then its energy represents the minimum
achievable value. To make a link with the XY model, we could say that factorization point for the Heisenberg model corresponds to $h=0$.

In the absence of spontaneous symmetry breaking, i.e. for finite systems,
and in the absence of external fields, the ground state belongs to an $\left(
N+1\right) $-dimensional manifold, and can be expanded in the
over-complete set of factorized states
\begin{equation}
\left| \Phi \right\rangle =\frac{1}{\mathcal{N}^{\prime }}\int d\theta \int
d\phi f\left( \theta ,\phi \right) \left| \Phi \left( \theta ,\phi \right)
\right\rangle ,
\end{equation}
where $f\left( \theta ,\phi \right) $ is a weight function.The inner product
between states pointing in different directions reads
\begin{equation}
\left\langle \Phi \left( \theta ^{\prime },\phi ^{\prime }\right) |\Phi
\left( \theta ,\phi \right) \right\rangle =\left[ \cos \theta \cos \theta
^{\prime }+e^{i\left( \phi -\phi ^{\prime }\right) }\sin \theta \sin \theta
^{\prime }\right] ^{N}.  \label{inner}
\end{equation}
Then, only in the thermodynamic limit we have a set of orthogonal states.
Given the continuous $SU(2)$ symmetry of the model, spontaneous symmetry
breaking implies that the system will select one direction out of all the possible
choices in the $\left( \theta ,\phi \right) $ space.

As we are interested in studying the problem of vanishing of
peculiar quantum properties, we shall choose initial states with
given symmetry properties, which, in the finite size limit, do
exhibit those properties.

\section{Transition to the Schr\"{o}dinger cat regime\label{2sp}}

According to the superposition principle, every linear combination of
quantum states is allowed. On the other hand, it is well known that
superposition cannot be observed on the macroscopic scale because of
superselection, the most convincing argument being the Schr\"{o}dinger cat
paradox. Then, on this scale, all but a small set of states belonging the
total Hilbert space are actually forbidden. This process, which leads to a
diagonal form of the density operator in a preferred basis, implies the
vanishing of the most peculiar of quantum properties: state interference.

In order to study the vanishing of state interference we analyze two
different quantities: two-spin entanglement and the fluctuation properties
of the order parameter $M_{x}=\left( \sum_{l}\sigma _{l}^{x}\right) /N$.
Given a density matrix $\rho $, fluctuations statistics %of a generic
%quantity $\hat{A}$ ({\bf chi \`e questo $\hat{A}$ nel seguito?})
 is associated to the generating function
\begin{equation}
G_{\rho }\left( \lambda \right) =\mathrm{Tr}\left\{ \rho e^{\frac{i\lambda }{%
N}\sum_{l}\sigma _{l}^{x}}\right\} ,
\end{equation}
which is the Fourier transform of the probability distribution
function of $M_{x}$. When the system can be observed in $m$ states
$\Psi _{1},\Psi _{2},\ldots \Psi _{m}$, whit related generating
functions $G_{\Psi _{n}}\left( \lambda \right)$, quantum
superposition effects appear if
\begin{equation}
\Delta G=G_{\rho }\left( \lambda \right) -\frac{1}{m}\sum_{n=1}^{m}G_{\Psi
_{n}}\left( \lambda \right) \neq 0.
\end{equation}
We expect that in the symmetry broken regime $\lim_{N\rightarrow
\infty }\Delta G=0$.

Similar considerations, carried out about entanglement properties,
lead to establish that, when $N\rightarrow \infty$, only
factorized states can be observed.

Even if superselection can be assumed as a principle, the size dependence of
quantum interference effects will be related to the particular system
observed. In the following, we find different decaying behaviors for
the XY and the XXX models, which are caused by the difference in the symmetry
of the two systems. We will start in both models by considering symmetric
states (which are expected  not to survive in the  thermodynamic limit) and
we shall study two-spin entanglement as a function of $N$. In fact, the existence of
degenerate ground state manifolds for any $N$ allows to calculate coherence
properties as a function of the size of the system.

For qubit systems, like spins, two-body entanglement can be measured through
concurrence \cite{wootters}. As shown in Ref. \cite{verrucchi}, for states
which are invariant under the action of the parity operator, the concurrence
$\mathcal{C}_{ij}$ of two spins at sites $i$ and $j$ is related to the
quantum correlation functions by simple relations which will be used here.
As pointed out in Ref. \cite{rossignoli}, for the XY model at the factorizing
field the two-spin concurrence does not depend on the spin distance $|i-j|$. Similar
arguments can be used also for the XXX chain. Since we are dealing with
superpositions of ferromagnetic states, the entanglement will be
of ferromagnetic kind as well. In this case,
\begin{equation}
\mathcal{C}_{ij}=\frac{1}{2}\left| p_{1}-p_{2}\right| -p_{III},
\end{equation}
where $\left| p_{1}-p_{2}\right| $ is the average value of $\left| \uparrow
\uparrow \right\rangle \left\langle \downarrow \downarrow \right| +\left|
\downarrow \downarrow \right\rangle \left\langle \uparrow \uparrow \right| $%
, and $p_{III}$ is the average value of $\left| \uparrow \downarrow
\right\rangle \left\langle \uparrow \downarrow \right| $.

\subsection{XY model}

Let us first consider the order parameter fluctuations for the symmetric states $\left| \alpha ^{\pm }\right\rangle $. These states could be obtained by starting with $h\neq h_F$. In this case the exact ground state would have definite parity. For instance, for $h> h_F$, the ground state is even. By lowering the field until the value $h_F$ is reached, the system is driven in  $\left| \alpha ^{+ }\right\rangle $.The generating function
is
\begin{equation}
G\left( \lambda ,\alpha ^{\pm }\right) =\frac{1}{4u_{\pm }^{2}}\left[
G\left( \lambda ,\Psi _{F}^{+}\right) +G\left( \lambda ,\Psi _{F}^{-}\right)
+\tilde{G}\left( \lambda ,\Psi _{F}^{+},\Psi _{F}^{-}\right) +\tilde{G}%
\left( \lambda ,\Psi _{F}^{-},\Psi _{F}^{+}\right) \right],
\end{equation}
where
\begin{equation}
\tilde{G}\left( \lambda ,\Psi _{F}^{\pm },\Psi _{F}^{\mp }\right)
=\left\langle \Psi _{F}^{\pm }\right| e^{\frac{i\lambda
}{N}\sum_{l}\sigma _{l}^{x}}\left| \Psi _{F}^{\mp }\right\rangle.
\end{equation}
It is easy to show that
\begin{equation}
G\left( \lambda ,\Psi _{F}^{\pm }\right) =\left( \cos \frac{\lambda }{N}%
\pm i\sin \frac{\lambda }{N}\sin 2\theta \right) ^{N},
\end{equation}
\begin{equation}
\tilde{G}\left( \lambda ,\Psi _{F}^{\pm },\Psi _{F}^{\mp
}\right)=\left( \cos \frac{\lambda }{N}\cos 2\theta \right) ^{N}.
\end{equation}
Then, interference effects (manifested by the off diagonal
elements) disappear exponentially with $N$.

As a second characterization, we study the concurrence for the symmetric
states $\left| \alpha ^{\pm }\right\rangle $. This can be easily derived
using the expression of $\left| \alpha^{\pm }\right\rangle $ in terms of $\left| \Psi _{F}^{\pm }\right\rangle $.
The result (see also Ref. \cite{rossignoli}) is given by
\begin{equation}
\mathcal{C}_{ij}\left( \alpha ^{\pm }\right) =\left( \cos 2\alpha \right)
^{N-2}\frac{\sin ^{2}2\alpha }{1\pm \left( \cos 2\alpha \right) ^{N}},
\end{equation}
where the factor $\left( \cos 2\theta \right) ^{N-2}$ derives from the
non-orthogonality of $\left| \Psi _{F}^{+}\right\rangle $ and $\left| \Psi
_{F}^{-}\right\rangle $ and determines the speed of classicalization. In the
macroscopic limit, $\mathcal{C}_{ij}\left( \alpha ^{\pm }\right) $ vanishes
as $\left[ \left( 1-\gamma \right) /\left( 1+\gamma \right) \right] ^{N/2}$.
Then, for every finite value of the anisotropy $\gamma $, entanglement
decays exponentially with $N$. A small but finite anisotropy will enhance
entanglement. Actually, the $\gamma =0$ limit implies a non-analytic change
% in the model, which would assume a continuous symmetry. In Fig. \ref{Fig1}
% we plot $\mathcal{C}_{ij}\left( \alpha ^{\pm }\right) $ for different values
% of the anisotropy.

\subsection{XXX model}

In analogy with the previous case, we introduce a state which is
invariant under a given spin rotation. In particular, if we choose
the state $\left| \Phi _{e}\right\rangle$ invariant under
rotations about the $y$ axis  $\exp \left[ -i\theta \sum_{l}\sigma
_{l}^{y}\right] \left| \Phi _{e}\right\rangle =\left| \Phi
_{e}\right\rangle $, we have
\begin{equation}
\left| \Phi _{e}\right\rangle =\frac{1}{\mathcal{N}}\int_{0}^{2\pi
}d\theta \left| \Phi _{\theta }\right\rangle,   \label{equat}
\end{equation}
where $\left| \Phi _{\theta }\right\rangle =\otimes _{l}\left| \Phi
_{l}\left( \theta ,0\right) \right\rangle $, and where $\left| \Phi
_{l}\left( \theta ,0\right) \right\rangle =\cos \theta \left| \uparrow
_{l}\right\rangle +\sin \theta \left| \downarrow _{l}\right\rangle $.
Requiring the normalization of $\left| \Phi _{e}\right\rangle $ implies $%
\int d\theta ^{\prime }d\theta \left[ \cos \left( \theta -\theta ^{\prime
}\right) \right] ^{N}=\mathcal{N}^{2}$. It is easy to verify that $\left|
\Phi _{e}\right\rangle $ is also an eigenstate of the parity operator. In fact, the
integration over $\theta $ cancels, in the superposition, all terms with an odd number of down spins. Each $\left| \Phi _{\theta }\right\rangle$ would be the actual ground state in the presence of an external field directed along the direction $\theta $.

Let us analyze the order parameter. First, we calculate fluctuations for a
given element of the ground state degeneracy manifold, obtaining
\begin{equation}
G_{\Phi _{\theta }}\left( \lambda \right) =\frac{1}{\mathcal{N}^{2}}\left(
\cos \frac{\lambda }{N}+i\sin \frac{\lambda }{N}\sin 2\theta \right) ^{N}.
\end{equation}
For large $N$ we see that $G_{\Phi _{\theta }}\left( \lambda \right) \simeq
\exp \left( i\lambda \sin 2\theta \right) $. \ This is the typical
expression of the generating function of a non fluctuating quantity. Its
Fourier transform, which is the probability distribution function of the
order parameter is indeed, for any $\theta $, a Dirac's delta distribution
around $\sin 2\theta $.

Furthermore, in the superposition state we have
\begin{equation}
G_{\Phi _{e}}\left( \lambda \right) =\frac{1}{\mathcal{N}^{2}}\int d\theta
^{\prime }d\theta \left[ g\left( \lambda ,\theta ,\theta ^{\prime }\right) %
\right] ^{N},
\end{equation}
where
\begin{equation}
g\left( \lambda ,\theta ,\theta ^{\prime }\right) =\cos \frac{\lambda }{N}%
\cos \left( \theta -\theta ^{\prime }\right) +i\sin \frac{\lambda }{N}\sin
\left( \theta +\theta ^{\prime }\right).
\end{equation}
As $N$ gets large, the vanishing of interference is observed. In the large $N$
limit, the steepest descent method gives
\begin{equation}
\Delta G_{\Phi _{e}}\left( \lambda \right) =\int d\theta G_{\Phi _{\theta
}}\left( \lambda \right) \left[ \exp \left( \frac{\lambda ^{2}\cos 2\theta }{%
2N}\right) -1\right],
\end{equation}
implying
\begin{equation}
\Delta G_{\Phi _{e}}\left( \lambda \right) \sim 1/N
\end{equation}
in the asymptotic regime.

As far as the two-spin concurrence is concerned, it is straightforward to find
\begin{equation}
\left| p_{1}-p_{2}\right| =\frac{1}{\mathcal{N}^{2}}\int d\theta
d\theta ^{\prime }\left[ \cos ^{2}\theta \sin ^{2}\theta ^{\prime
}+\cos ^{2}\theta ^{\prime }\sin ^{2}\theta \right] \left[ \cos
\left( \theta -\theta ^{\prime }\right) \right] ^{N-2},
\end{equation}
and
\begin{equation}
p_{III}=\int d\theta d\theta ^{\prime }\cos \theta \cos \theta
^{\prime }\sin \theta \sin \theta ^{\prime }\left[ \cos \left(
\theta -\theta ^{\prime }\right) \right] ^{N-2},
\end{equation}
yielding
\begin{equation}
\mathcal{C}_{ij}\left( \Phi _{e}\right) =\frac{1}{2}\frac{\int d\theta
d\theta ^{\prime }\left[ \tan \left( \theta -\theta ^{\prime }\right) \right]
^{2}\left[ \cos \left( \theta -\theta ^{\prime }\right) \right] ^{N}}{\int
d\theta d\theta ^{\prime }\left[ \cos \left( \theta -\theta ^{\prime
}\right) \right] ^{N}}.
\end{equation}
This result is very simple for $N$ even. In that case one gets
\begin{equation}
\mathcal{C}_{ij}^{N\text{\textrm{\ even}}}\left( \Phi _{e}\right) =\frac{1}{%
2\left( N-1\right) }.
\end{equation}
This result can be understood taking into account that, given Eq. (\ref
{inner}), the inner product between $\left| \Phi \left( \theta ,0\right)
\right\rangle $ and $\left| \Phi \left( \theta ^{\prime },0\right)
\right\rangle $ vanishes exponentially with $N$ . This allows to evaluate
integrals, in the large $N$ regime, by means of the steepest descent method.
It is clear that, when $N$ gets large, $\left[ \cos \left( \theta -\theta ^{\prime
}\right) \right] ^{N}$ is different from zero only for $%
\left( \theta -\theta ^{\prime }\right) $ $\simeq 0$. Expanding all
terms around this value, the concurrence is well approximated by the ratio
between two Gaussian integrals
\begin{equation}
\mathcal{C}_{ij}\left( \Phi _{e}\right) \simeq \frac{1}{2}\frac{%
\int_{-\infty }^{\infty }x^{2}\exp \left( -\frac{Nx^{2}}{2}\right) dx}{%
\int_{-\infty }^{\infty }\exp \left( -\frac{Nx^{2}}{2}\right) dx},
\end{equation}
which eventually gives $\mathcal{C}_{ij}\left( \Phi _{e}\right) \simeq 1/2N$
in the asymptotic regime.

\section{Discussion\label{disc}}

\begin{figure}%[b]
  \includegraphics[width=.5\textwidth,height=40mm]{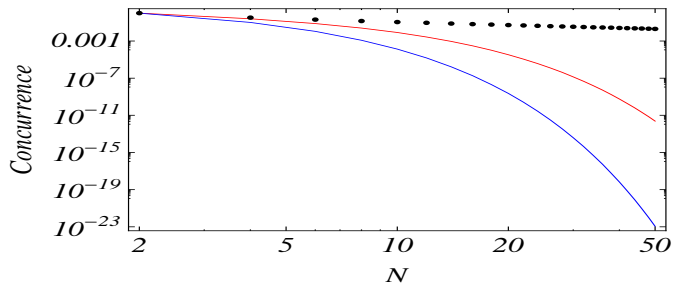}
\caption{Two-spin concurrence as a function of
   the system size $N$ for the XXX model (dots) and for the XY model (continuous lines).
    In this latter case, we considered two different anisotropy parameters: the red line represents $\gamma=0.5$, and the blue line
      corresponds to $\gamma=0.8$.}
\label{figura}
\end{figure}

We tackled the problem of describing how quantum coherence effects
vanish as the system size becomes macroscopic. Even if this
phenomenon is expected to appear in generic systems, two different
symmetry-broken model have been considered when exact and analytic
treatment are possible. In the first one (the XY model in a
transverse field), because of the discrete symmetry, the ground
state spans a two-dimensional manifold. As for the Heisenberg
model, the dimension of the manifold grows with $N$, eventually
reaching a dense structure. In Fig. \ref{figura} we plot the
behavior of the concurrence $\mathcal{C}_{ij}$ for the two models.
In the case of the XY chain, we used also different values of the
anisotropy
.
% \begin{figure}
%   % Requires \usepackage{graphicx}
%   \includegraphics{c}\\
%   \caption{Two-spin concurrence as a function of
%    the system size $N$ for the XXX model (dots) and for the XY model (continuous lines).
%     In this latter case, we considered different anisotropy parameters: the red line stands
%      for $\gamma=0.1$, the blue line represents $\gamma=0.5$, and the green line
%       corresponds to $\gamma=0.8$.  }\label{figura}
% \end{figure}
A necessary step to determine local quantities, like two-spin
entanglement or magnetization, is the introduction of the reduced
density matrix. Given the peculiar structure of the factorized
states we have considered here, calculating the reduced density
matrix requires the computation of inner products between states
defined on $\left( N-m\right) $ (where $m$ is a finite number)
spin subspaces) aligned along different directions . Both for the
XY and for the isotropic model these quantities vanish
exponentially in the large $N$ limit, destroying in such a way the
quantum interference between different components. However, since
the XXX model has a continuous symmetry, the ground state manifold
is continuous as well, and all matrix elements are integrated. The
integration implies a reduction in the decoherence rate, which
turns out to be linear in $1/N$.

% In conclusion, it the kind of symmetry broken by the thermodynamic
% limit influences deeply the disappearance of quantum superposition
% effects. The behavior of the order parameter fluctuations and of
% two-spin entanglement are a direct consequence of such
% classicalization process.

In this paper we considered two exactly solvable models, and studied how superselection tends to destroy their quantum properties. A typical tool used to study problems whose solution is not known is the mean-field approximation, that, in fact, consists in the introduction of ``product states'' with the same aspect of those described in this paper. For example, in the BCS theory of superconductivity the solution introduced is factorized in the space of the modes $k$. Since in the finite-size case the symmetry is expected to be conserved, while the mean-field states are widely unsymmetrical, the linear superposition of degenerate states is a way to restore it. Once the thermodynamic limit is performed, all the considerations made in this paper apply. Then, we can conclude that our results apply not only to the models explicitly studied, but they could used, in the limit of validity of the mean-field theory, in all systems belonging to the classes of symmetry discussed.

\acknowledgments
  The authors wish to acknowledge S. Paganelli for useful discussions. GLG acknowledges the “Juan de la Cierva” fellowship of the Spanish Ministry of Science and Innovation.

\end{document}